\documentclass[runningheads]{llncs}
\usepackage{tabu}
\usepackage{tikz}
\usepackage{listings}
\usepackage{xcolor}
\usepackage[normalem]{ulem}
\usepackage{wrapfig}
\usepackage{amssymb}
\usepackage{amsmath}
\usepackage{xspace}
\usepackage{hyperref}
\colorlet{brightlightbluegray}{red!9!green!8!blue!6}%
\colorlet{redgb}{red!91!green!92!blue!94}%
\colorlet{rgreenb}{green!70!black}%
\colorlet{rgblue}{red!20!green!20!blue!94}%
\colorlet{mediumbluegray}{blue!20!gray!40}%
\usetikzlibrary{arrows.meta, positioning, shapes.geometric}

\newcommand{\fsl}{\textsl}
\newcommand{\fsf}[1]{{\footnotesize{\textsf{#1}}}}

\newcommand{\mname}[1]{\fsl{#1}}
\newcommand{\pname}[1]{\fsl{#1}}
\newcommand{\rname}[1]{\textsc{#1}}
\newcommand{\paraname}[1]{\fsf{#1}}
\newcommand{\val}[1]{\texttt{#1}}

\usepackage{url}
\newcommand{\mps}[1]{}
\newcommand{\akc}[1]{}
\newcommand{\oj}[1]{}
\newcommand{\hide}[1]{}

\newcommand{\ahoy}{Ahoy\xspace}

\newcommand{\msf}{\mathsf}

\newcommand{\inn}{\ensuremath{\ulcorner\msf{in}\urcorner\,}}
\newcommand{\out}{\ensuremath{\ulcorner\msf{out}\urcorner\,}}
\newcommand{\nil}{\ensuremath{\ulcorner\msf{nil}\urcorner\,}}

\usepackage[inline]{enumitem}

\begin{document}
\title{\ahoy: LLMs Enacting Multiagent Interaction Protocols}
\titlerunning{\ahoy}

\author{Omkar J. Joshi\inst{1}\orcidID{0009-0006-5069-2326} \and
Munindar P. Singh\inst{1}\orcidID{0000-0003-3599-3893} \and
Amit K. Chopra\inst{2}\orcidID{0000-0003-4629-7594} }
\authorrunning{O. Joshi et al.}

\institute{North Carolina State University, Raleigh, NC, USA \and
Lancaster University, Lancaster, UK }

\maketitle

\begin{abstract}
An interaction protocol formalizes how the agents in a multiagent system interact, which facilitates implementing agents.
Existing approaches yield agent implementations specific to the selected protocols. 
\emph{How can we engineer intelligent agents that can enact protocols but are programming-free?}
Our contribution, \emph{\ahoy}, addresses this question by creating LLM agents that dynamically select and enact declarative protocols to achieve user goals. 
We demonstrate that an \ahoy agent can correctly and intelligently enact multiple protocols---concurrently if appropriate to the user goal---without specialized training. 
\ahoy's significance lies in that it brings together declarative protocols and LLMs, both approaches that promise improved knowledge engineering for agents. 
\end{abstract}

\keywords{Agentic AI, Communication Protocols, Information Protocols}

\section{Introduction}
\label{Introduction}
A sociotechnical system (STS) is a system of autonomous real-world principals interacting with each other and sharing information and other resources \cite{chopra:iose:2016}.
STSs apply in e-commerce, health, and finance. 
An STS can be realized as a multiagent system: each agent represents and interacts on behalf of its principal with other agents; no agent is subordinate to a central controller.

An \emph{interaction protocol} operationalizes decentralization by specifying the interaction constraints each agent must respect.
Recent work on protocols emphasizes declarative, information-based approaches exemplified by the \emph{Blindingly Simple Protocol Language} (BSPL) \cite{singh:bspl:2011}. 
The declarative approaches offer notable advantages over communicating state machine-based approaches \cite{chopra:langeval:2020}, including supporting asynchronous, flexible interactions between agents and enabling higher-level meaning through formal properties like commitments \cite{baldoni:2CL:2014,winikoff:commitments:2007,yolum:flexible-protocols:2002}.
Being formal, declarative protocols are amenable to model checking \cite{chopra:mambo:2025,singh:tango:2021}: thus, a protocol may be verified for correctness before it is adopted. 

A programming model facilitates agent implementation.
BSPL supports programming models that structure an agent's internal reasoning based on communications sent and received in light of a protocol.
Kiko \cite{christie:kiko:2023}, one such programming model, facilitates engineering Python agents; others facilitate engineering \emph{Belief-Desire-Intention} agents \cite{baldoni-aaai:orpheus:2025,chopra:azorus:2025}.  
These programming models facilitate building and maintaining agents that are programmed to enact the relevant protocols.  The programming effort (e.g., Listing~\ref{lst:kiko-pattern}) concerns encoding an agent's reasoning to take actions made available by a protocol.  In contrast, we ask this question: \emph{How can we create agents that without protocol-specific programming effort?}

Agentic AI applies Large Language Models (LLMs) as the reasoning engines of agents.  
Agentic AI aims to reduce the knowledge engineering effort: rather than manually encoding world knowledge, it leverages the extensive knowledge LLMs have learned from training data.
Indeed, LLMs have demonstrated significant ability at diverse tasks \cite{li-llm-domain-survey-2024,naveed-comprehensive-llm-overview-2025}. Thus, we refine our question to this: \emph{How can LLM agents enact declarative protocols?} 
If LLM agents could enact interaction protocols, the resulting implementation would be programming-free: LLMs would serve as the agent reasoning substrate.   

Accordingly, we contribute \ahoy, an LLM-based approach for creating agents that select and enact BSPL protocols using Kiko.
An \ahoy agent takes a user's goals in natural language.
The Kiko adapter maintains the states of ongoing protocol enactments and engages the LLM to reason about protocol actions.
Based on the user goal and the current state of the agent's enactments, the \ahoy agent sends messages from the relevant protocols using the Kiko adapter to progress towards the user goal.  
We demonstrate that an \ahoy agent can adopt and enact roles in multiple protocols---sequentially or concurrently---without additional programming. 
It intelligently enacts flexible protocols and respects protocol constraints.
An \ahoy agent can handle external events while doing so, which simplifies interoperation with real-world \emph{event source} by requiring only minor changes to accommodate specific event formats.

\section{Background}
\label{sec:Background}

\subsection{Specifying Interaction Protocols}
BSPL protocols specify the interaction between agents using roles, messages, and parameters.
Each message (schema) specifies its sender and receiver roles and its parameters.
Message schemas are constrained by \emph{parameter adornments} that enforce information causality: agents can only send messages when they have the knowledge required by those messages' adornments.
An agent's \emph{local state} includes the history of the role the agent is playing and the set of bindings established thus far.
In BSPL, an \emph{enactment} is a complete assignment of bindings to all parameters in all messages, satisfying all adornment constraints.
\emph{Key Parameters} uniquely identify an enactment, ensuring that at most one instance of an enactment occurs per unique key binding.
Listing~\ref{lst:bspl-spec} shows the \pname{Purchase} protocol, which we use to illustrate these concepts.

\begin{lstlisting}[caption={The \mname{Purchase} protocol in BSPL.}, label=lst:bspl-spec]
Purchase {
  roles Buyer, Seller, Shipper
  parameters out ID key, out item, out price, out outcome
  private address, resp, shipped, satisfaction

  Buyer -> Seller: rfq[out ID,out item]
  Seller -> Buyer: quote[in ID, in item, out price]

  Buyer -> Seller: accept[in ID, in item, in price, out address, out resp]
  Buyer -> Seller: reject[in ID, in item, in price, out outcome, out resp]

  Seller -> Shipper: ship[in ID, in item, in address, out shipped]
  Shipper -> Buyer: deliver[in ID, in item, in address, out outcome]

  Buyer -> Seller: completed[in ID, in item, in price, out satisfaction]
}
\end{lstlisting}

\pname{Purchase} defines three roles: \rname{buyer}, \rname{seller}, and \rname{shipper}.
It declares global parameters (\paraname{ID}, \paraname{item}, \paraname{price}, \paraname{outcome}).
The \emph{private} line declares local parameters (\paraname{address}, \paraname{resp}, \paraname{shipped}, \paraname{satisfaction}) that are used by internal messages (schemas) used by specific roles.
These parameters remain hidden from the protocol's public interface to encapsulate details irrelevant to the overall interaction.
This separation enables BSPL to manage enactment uniqueness while supporting modular protocol composition.

\emph{Parameter Adornments:}
Parameter adornments capture the viability of message emissions based on an agent's knowledge.

The \inn~adornment marks a parameter whose binding the agent \emph{must already know}.
For example, for the \mname{quote} message in \pname{Purchase}, \rname{seller} must already know the value bound to \inn~\paraname{ID} and \inn~\paraname{item} (received in the prior \mname{RFQ} message) before sending \mname{quote}.
A message instance can only be sent by an agent if all its \inn parameters are bound in the agent's local state.

Conversely, the \out~adornment marks a parameter whose binding the agent \emph{generates when sending} the message instance---these must be new bindings that did not previously exist in the agent's local state.
For the \mname{RFQ} message in \pname{Purchase}, \rname{buyer} generates a unique \out~\paraname{ID} and selects an \out~\paraname{item}.
Each \out~parameter has exactly one binding per enactment which remains immutable throughout that enactment.

Finally, the \nil~adornment marks a parameter whose binding the agent \emph{must not know}.
An agent can only send a message instance when all its \nil~parameters are unbound in the agent's local state.
The \nil~adornment enables conditional message flows: certain messages become unavailable once specific bindings are established, creating mutually exclusive enactment paths.

\subsection{Implementing Protocol-Based Agents}
Kiko is a programming model for implementing BSPL agents in Python.
It abstracts an agent's communication service, enabling developers to focus on encoding the agent's internal reasoning.
To use the Kiko adapter, the developer needs to provide the multiagent system configuration and write one or more \emph{decision makers}.
A decision maker is a function that the Kiko adapter invokes in response to specific events, such as when a specific message is received or when information bindings change.
The developer is expected to interact with \emph{forms}, which are message templates provided by the adapter.
Each form corresponds to a potential message that the agent may send, constrained by protocol rules.
To send a message, the agent needs to bind the required \out parameters according to business logic implemented in the agent's decision maker function.
This transforms the form (template) into a concrete message instance which can be sent by the adapter.

\begin{lstlisting}[caption={Kiko sample showing the @adapter decorator.},label=lst:kiko-pattern]
@adapter.decision(event=InitEvent)
def start(enabled):
    for item in ["ball","bat"]:
        ID = str(uuid.uuid4())
        for m in enabled.messages(RFQ):
            m.bind(ID=ID,item = item)
\end{lstlisting}

Listing~\ref{lst:kiko-pattern} demonstrates this pattern through a Python function \texttt{start()} that serves as a decision maker for the \rname{Buyer} agent in the \pname{Purchase} protocol.
The function is registered using the \texttt{@adapter.decision} decorator and is triggered by an \texttt{InitEvent}.
When invoked, the adapter passes the set of enabled messages to the function.
The decision maker then iterates through this set to find the \texttt{RFQ} forms and uses the \texttt{.bind()} method to assign values to the \paraname{ID} and \paraname{item} parameters.
Upon the decision maker's completion, the adapter automatically collects the completed instances, validates them against the protocol's integrity constraints, and handles their emissions over the network.

\subsection{Programming LLM Agents}
Large Language Models (LLMs) can serve as the reasoning component in agent architectures, enabling agents to plan tasks and make decisions guided by natural language instructions \cite{brown-language-2020,ouyang-training-2022}.
In an LLM agent, the model is invoked repeatedly in a loop, each time conditioned on the user goal, the agent's current state, and the available actions.

Programming an LLM agent typically involves the following components:
A system prompt describes the reasoning strategy (e.g., deliberate step-by-step reasoning) \cite{wei-chain-thought-2022} and the agent's environment. 
In Ahoy, the \emph{system prompt} defines the agent's role, protocol constraints that the agent must satisfy (such as respecting message preconditions and parameter types) and any instructions that are unchanged across decisions.
Additionally, the system prompt may also contain stylistic guidelines or formatting instructions to improve response quality.
The \emph{user prompt} clarifies the task instance and the relevant, up-to-date context (e.g., dialog history and event description).
Because context windows are limited, the information in the system and user prompts should minimize redundancy, with all content structured for clarity.

LLM agents can perform tasks effectively when they can \emph{invoke tools} (e.g., web search, database access, and calculators) and incorporate tool outputs into subsequent reasoning.
Recent approaches show that LLMs can be prompted (or trained) to interleave reasoning and acting \cite{yao-react-2022}, and use external tools in a standardized way \cite{schick-toolformer-2023}.

To enable long-term tasks, agents should maintain and use a persistent memory of past decisions and outcomes.
A common approach is \emph{Retrieval Augmented Generation (RAG)}, where the agent retrieves relevant information to augment the next prompt \cite{lewis-rag-2020}.
In practice, memory combines structured state (e.g., key-value stores), historical summaries (e.g., trace compression), and retrieved knowledge from databases (e.g., vector embeddings).

At runtime, the LLM agent executes an observe-reason-act loop formed from these components.
Listing~\ref{lst:llm-agent-loop} in the appendix shows an example of this loop.

\section{Architecture}
\label{sec:architecture}

\begin{figure}[htb]
    \centering
    \includegraphics[width=\textwidth]{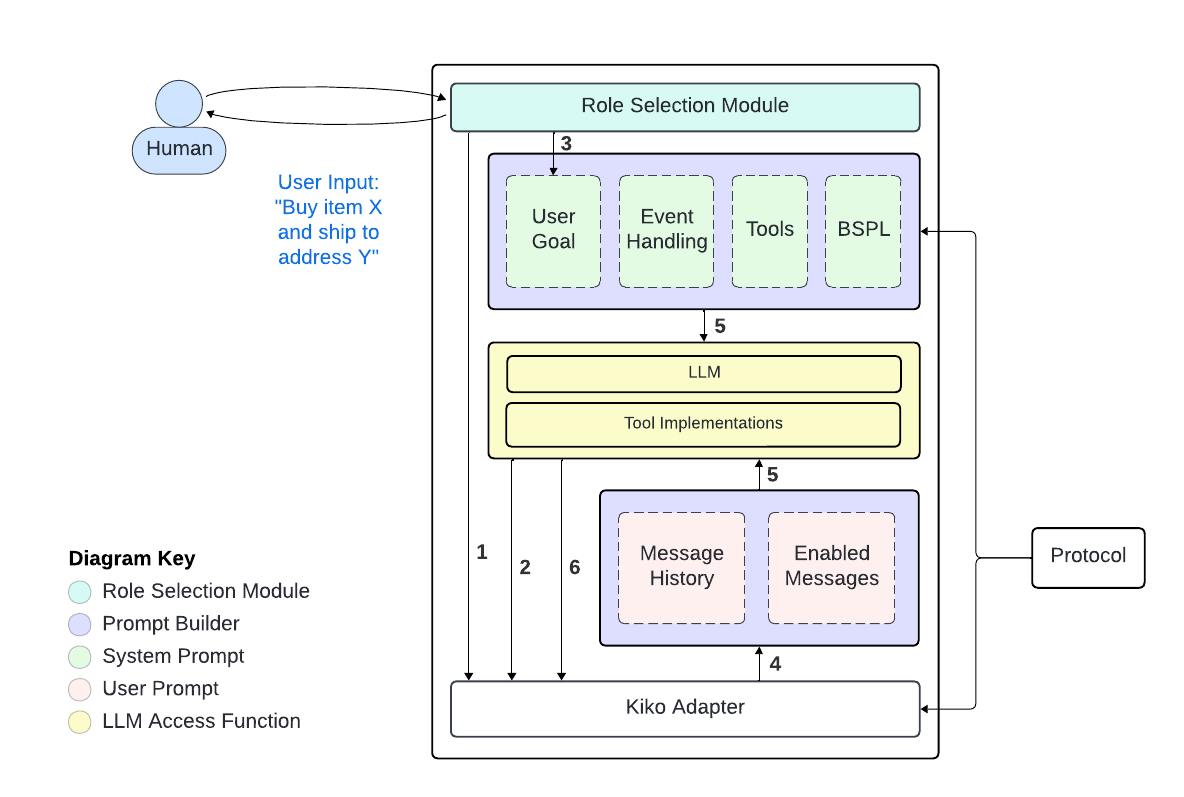}
    \caption{\ahoy architecture and main steps. (1) The Role Selection Module configures the Kiko adapter for the multiagent system. (2) \ahoy registers the LLM Access Function as an adapter callback. (3) The Role Selection Module maps user input to user goal. (4) Adapter detects event; Prompt Builder Module constructs user prompt by accessing the adapter. (5) The LLM Access Function calls the LLM with the system and user prompts received from the Prompt Builder Module. (6) Return selected message instances to the adapter.}
    \label{fig:architecture_diagram}
\end{figure}

We make an \ahoy agent programming-free by decoupling LLM reasoning from protocol-specific logic, as shown in Figure~\ref{fig:architecture_diagram}.

\begin{figure*}
    \centering
    \includegraphics[width=0.65\textwidth]{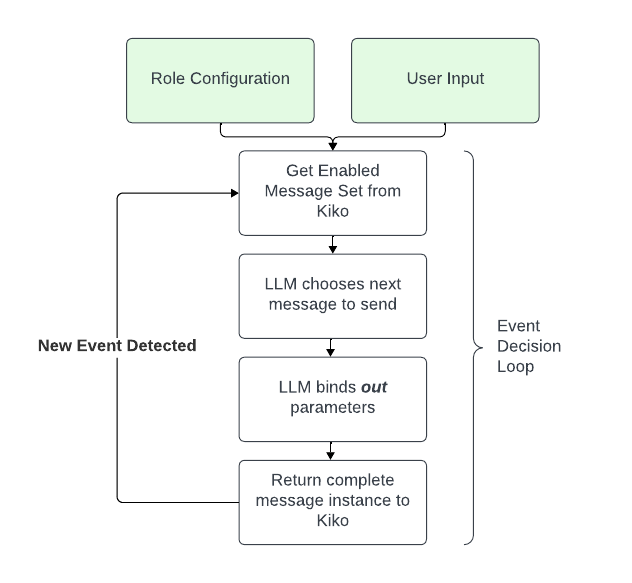}
    \caption{Protocol enactment using {\ahoy}.}
    \label{fig:timeline_diagram}
\end{figure*}

The \ahoy architecture prompts the LLM to reason over constraints and possibilities defined by BSPL protocols and user requirements.
This architecture consists of three modules that interact via a defined flow.
\begin{itemize}
    \item The \emph{Role Selection Module} enables users to select protocols and roles, processes user input, and initializes the Kiko adapter with the MAS configuration.
    \item The \emph{Prompt Builder} extracts the local state information and formats it into structured LLM input.
    \item The \emph{LLM Access Function}  processes tool invocations and calls the LLM to reason about which messages to send and how to bind their parameters. 
\end{itemize}
During a protocol enactment, as shown in Figure~\ref{fig:timeline_diagram}, \ahoy runs an event loop.
Initially, the user converses with the Role Selection Module in natural language to select the needed protocols, configures \ahoy with a set of roles from the selected protocols, and provides a goal for the agent to achieve.
The Role Selection Module instantiates the Kiko adapter for the selected protocol-role pairs and registers the LLM Access Function as a callback.
\ahoy then creates a system prompt containing the user goal, foundational knowledge of BSPL semantics, tool and event-handling guidance, and the selected protocols.
This process happens once per enactment.
During protocol enactment, the Kiko adapter monitors the agent's local state.
When a decision event occurs---whether from a message arrival, a change in the set of enabled messages, or an external event from a connected event source---the Kiko adapter calls the LLM using the registered callback.
The LLM selects the messages to send, binds their parameters, and uses the Kiko adapter to send them.
The individual modules are explained below.

\subsection{Role Selection Module}
This module presents a list of available protocols and roles to the user, who selects the roles for the agent to play during the current enactment.
An \ahoy agent can play multiple roles in multiple protocols at the same time without user guidance on how to implement them.
This module is responsible for configuring the Kiko adapter with the roles and protocols selected by the user.
To achieve this, the Role Selection Module uses the protocol specifications annotated with comments that explain the function of each message.
Examples of the annotated protocol specifications are included in the appendix.
Each role needs to be mapped to a \emph{termination condition}.
These termination conditions are protocol-defined constraints that specify when a role has finished participating in an enactment.
For example, the termination condition for \rname{Buyer} in the \pname{Purchase} protocol is sending \mname{completed}; while the termination condition for \rname{Merchant} in the \pname{Logistics} protocol is receiving \mname{packed}.
The Role Selection Module infers the termination conditions from the chosen protocols for the chosen roles.
Next, the user describes their goals via a statement of the form ``I want to buy an item X with a budget of \$Y, deliver it to address Z'' that does not need to contain any BSPL or Kiko-related details.
The Role Selection Module provides this user input to the Prompt Builder module.

\subsection{Prompt Builder}
\label{sub-sec:prompt_construction}
This module constructs the system and user prompts from user input, protocol state and external event information.
A system prompt is constructed once per protocol enactment.
A user prompt is constructed dynamically for each event.

\emph{System Prompt:}
We do not require that the models be pretrained on BSPL semantics.
The system prompt contains information that enables any LLM to understand and enact protocols specified in BSPL.
To do so, we provide the following information as part of the system prompt: (1) a description of BSPL semantics (roles, messages, and adornments) as shown in Listing~\ref{lst:protocol-string}, (2) the agent's assigned protocol roles, (3) available tools and their calling conventions, (4) event-handling instructions, (5) protocol specifications annotated with comments that describe each message, and (6) the user input.
We provide a full system prompt in Listing \ref{lst:system-prompt-instance} in the appendix.

\begin{lstlisting}[float=*,language=Python, caption={Description of BSPL Semantics.},label=lst:protocol-string]
BSPL defines multiagent protocols where agents play roles and coordinate via information causality.

PARAMETER ADORNMENTS (three types):
1. **in** (Causal): Must already know from prior messages. Information provided from previous messages in the protocol.
2. **out** (Generation): You generate this binding; it appears once per enactment, creating mutual exclusion. Your role generates unique values for instances.
3. **nil** (Negative): Must NOT know this binding. Used for mutually exclusive paths where an agent cannot act until certain information remains unknown.

Key parameters identify protocol instances. Messages are ordered by information flow according to causal dependencies.
\end{lstlisting}
 
\emph{User Prompt:}
For each decision event, \ahoy constructs a fresh user prompt by querying the Kiko adapter for information about the ongoing enactments.
The user prompt includes the message history for the role, the current set of enabled messages, and all parameters and their adornments.
This prompt encodes sufficient context for the LLM to make decisions while remaining concise enough to not exceed token limits.
A full user prompt is in Listing \ref{lst:user-prompt} in the appendix.

\subsection{LLM Access Function}
This module is an event handler that mediates between the Kiko adapter and the LLM.
It is triggered when a decision event is triggered by either an incoming message, a change in the set of enabled messages, or an external event (``Inventory out of stock alert'').
It selects which message to send next and determines what the appropriate parameter bindings should be.
When triggered, the module receives the system and user prompts from the Prompt Builder.
It then calls the LLM with these prompts.
The LLM evaluates the enabled messages and selects which messages to send and their parameter bindings.
The LLM Access Function parses the LLM output to extract the messages and parameter bindings, returning the selected message instances to the adapter.
Since the Kiko adapter ensures that preconditions are met, the LLM is prompted to reason only about domain logic.
The LLM is responsible for handling concerns such as appropriateness (``Is this the right price for this item?'') and goal alignment (``Does this item match the user's goals?'').
The LLM Access Function also executes tool invocations using the implementations provided by developers and returns their outputs to the LLM.
Upon receiving a tool invocation, the LLM Access Function executes the corresponding locally-implemented functions referenced by name on the user's machine and passes the results back to the LLM.

\section{Programming Model}

A key contribution of \ahoy is a programming model centered on the LLM Access Function.
This programming model can be used in three ways.

\subsection{Configuring \ahoy}
The user configures an \ahoy agent by selecting from a list of available roles through a chat interface.
The user describes the goals in natural language, as explained above.
The \ahoy agent then infers the termination conditions from the protocol at initialization and monitors them during the enactment.
The user does not need to possess any coding knowledge of protocols, adapters, or LLM functions to configure and use \ahoy.

\subsection{Adding a new protocol}
The user can define new protocols and configure agents for them using \ahoy.
This also requires no programming knowledge---the declarative BSPL syntax is sufficient.
The user simply needs to specify a protocol and add the appropriate multiagent system configuration for the Kiko adapter.

\subsection{Extending \ahoy}
Developers can extend \ahoy in three ways: (1) adding new tools, (2) implementing custom decision strategies that do not need LLM decision-making, or (3) connecting external event sources.
To modify \ahoy, developers need to understand the LLM Access Function pattern shown in Listing~\ref{lst:ahoy-dec}.

\begin{lstlisting}[language=Python, caption={Ahoy LLM Access Function Pattern},label=lst:ahoy-dec]
from bspl.adapter import Adapter
from lib.state_manager import extract_social_state
from lib.llm_client import AnthropicLLMClient, choose_and_bind

adapter = Adapter(Seller, systems, agents)
llm_client = AnthropicLLMClient()

async def llm_access_function(enabled_store, event):
    
    # OBSERVE: Extract current protocol state
    state = extract_social_state(adapter)
    enabled_messages = list(enabled_store.messages())
    
    # REASON: Delegate to LLM with full protocol context
    message_instance = await choose_and_bind(
        adapter=adapter,
        enabled_store=enabled_store,
        event=event,
        client=llm_client,
        timeout=30.0,
        current_protocol="Purchase",
        current_role="Seller"
    )
    
    # ACT: Return selected message for adapter to send
    # (Adapter enforces all constraint validation)
    return message_instance

# Register as adapter decision maker
@adapter.decision()
async def handle_decision(enabled_store, event):
return await llm_access_function(enabled_store, event)
\end{lstlisting}

The \texttt{extract\_social\_state()} function provides the complete local state extracted from the Kiko adapter.
This can be used to access message history and identify bound parameters.
Developers can supplement this with data from external event sources to inform the LLM's decision.
For example, if an \ahoy agent is enacting \pname{Purchase} as a \rname{Buyer}, the developer can call a function that returns the inventory, which can be used to decide which item to purchase next.
The \texttt{enabled\_store.messages()} structure contains a list of enabled messages managed by the Kiko adapter.
This is the set of choices available to the LLM in terms of interacting with other agents.
The \texttt{choose\_and\_bind()} function is used to communicate with the LLM.
It calls the prompt construction functions internally.
The constructed prompt is then used to call the LLM.
By passing the client as a parameter to \texttt{choose\_and\_bind()}, we ensure that this approach is not tied to a specific LLM API and can be ported across different vendors (e.g., Anthropic, OpenAI, HuggingFace).
Once \texttt{choose\_and\_bind()} returns a bound message instance, the Kiko adapter validates the message against protocol constraints and sends it to the appropriate recipient.
To connect our LLM Access Function with the Kiko adapter, we register it as an event handler following the pattern shown in Listing~\ref{lst:kiko-pattern}.

\begin{lstlisting}[caption={Tool invocation in Ahoy.},label=lst:ahoy-tools]
# LLM response can include tool requests for complex reasoning
response = await choose_and_bind(...)
# Tool requests are structured as:
# {"choice": null, "params": {}, "tool_requests": [
#   {"tool": "save_state_to_memory", 
#    "args": {"agent_name": "Seller", "key": "price_strategy", 
#             "value": "accept above $50"}},
#    ]}

# Ahoy executes requested tools
if tool_requests:
    for tool_req in tool_requests:
        tool_name = tool_req.get("tool")
        tool_args = tool_req.get("args", {})
        
        # Execute tool
        result = await execute_tool_call(tool_name, tool_args)
\end{lstlisting}

The system prompt specifies which tools are available to the \ahoy agent and when each tool may be evoked.
\ahoy's LLM Access Function executes these tools locally.
This requires the developer to implement local tools (e.g., save-to-memory functions).
Local execution mitigates security risks from third-party LLM APIs that have direct execution capabilities.
Once the local tool implementations are added to the tool library, the developer can enable LLM access to them by modifying the LLM Access Function following the pattern shown in Listing~\ref{lst:ahoy-tools}.
\ahoy supports iterative reasoning by feeding tool results back to the LLM in subsequent calls.

\section{Evaluation}
We evaluate \ahoy through a series of controlled enactments designed to assess these capabilities (all using Anthropic's Claude Haiku 4.5):
\begin{description}
    \item [Programming Freeness:] The same agent can enact different protocols correctly without code modification.
    \item [Concurrent participation in multiple protocols:] The same agent can enact multiple protocols as different roles without exceptions or interference.
    \item [Intelligent path selection:] The agent can correctly enact flexible protocols with branching based on user input.
    \item [Handling external events:] The agent can handle events from the Kiko adapter and external sources simultaneously during enactments.
    \item [Preservation of BSPL constraints via adapter enforcement:] Protocol constraints are maintained in all enactments.
\end{description}

With the caveat that this claim is limited to the evaluated scenarios, we find that across all enactments reported in this section:
\begin{enumerate*}[label=(\arabic*)]
    \item No malformed message instances were emitted;
    \item No BSPL adornment violations occurred;
    \item No adapter level constraint exceptions were triggered; and, 
    \item No proposed message was rejected by the Kiko adapter
\end{enumerate*}

\subsection{Programming Freeness}
\label{subsec:demo1}
We evaluate whether an \ahoy agent can enact different BSPL protocols without code changes or re-initialization.
We first configure the \ahoy agent to enact the \pname{Purchase} protocol as the \rname{Buyer} role.
Using the Role Selection Module, we select the protocol and the role, then specify the item, budget, and delivery address in natural language.
No modification is made to the LLM Access Function, prompt construction logic, or tool infrastructure.
The enactment is logged and analyzed.
We then configure the \ahoy agent to enact the \pname{Logistics} protocol playing the \rname{Merchant} role.
We make the same configuration change, but now specify the orders to be packed and the delivery addresses instead.

\begin{itemize}
\item \pname{Purchase}: The agent sent a \mname{RFQ} (Request For Quote) message instance for a pen with the delivery constraints specified, received quotes from competing \rname{seller}s (\$5 and \$4), and accepted the lowest price. Enactment time: 20 seconds, 4 LLM calls.
\item \pname{Logistics}: The agent coordinated label requests and wrapping with implicit recognition of fragile items. Enactment time: 63.55 seconds, 11 LLM calls.
\end{itemize}

Both protocols were enacted successfully.
The agent displayed context-aware behavior (evaluating the options and making the more economical choice as the \rname{Buyer}, coordinating multiple orders with labeling and wrapping as the \rname{Merchant}), with no specific guidance beyond the provided protocol.

\subsection{Concurrent Participation in Multiple Protocols}
\label{subsec:demo2}
We evaluate whether a single \ahoy agent can simultaneously enact multiple protocols while making coherent decisions across distinct roles.
We configure one \ahoy agent to play two roles concurrently: \pname{Purchase}:\rname{Buyer} and \pname{Logistics}:\rname{Merchant}.
The agent receives decision events from both protocols in an interleaved fashion.
We verify that \ahoy tracks message history independently for each protocol and that the LLM correctly infers termination conditions  (e.g., \rname{Buyer} must send \mname{completed}).  
The \ahoy agent transitions from \pname{Purchase} (sending three \mname{RFQ}s) to \pname{Logistics} (requesting labels for purchased items), then back to \pname{Purchase} (receiving quotes, accepting offers, receiving deliveries), and finally to \pname{Logistics} again (requesting wrapping and packaging for purchased items).
These transitions are sensible, respecting commonsense reasoning.
For example, because the user input specifies all required details, the \ahoy agent sends label requests and RFQs interleaved.
After purchasing the glass vase (\$42), ceramic plate (\$27), and wooden bat (\$39), the agent initiates logistics operations: it sends \mname{RequestLabel} for warehouse destinations and \mname{RequestWrapping} for individual items.
Both role terminations occur independently without blocking each other.
Throughout the 20-message trace, no parameter values leak between protocols and the message history is built correctly at each decision point.
We observe the following by inspecting the logs:

\begin{itemize}
    \item \pname{Purchase} binds three \paraname{ID}s (\{\val{ab534279\ldots, 0fcd009e\ldots, ebbadddb\ldots}\}); \pname{Logistics} binds two \paraname{orderID}s (\{\val{32dc1cf9\ldots, 6aa820dd\ldots}\}). The parameter bindings from these enactments do not affect each other.
    \item \pname{Purchase}:\rname{Buyer} completes by sending \mname{completed}(\paraname{ID}=\val{ebbadddb-\ldots}); \pname{Logistics}:\rname{Merchant} completes by receiving \mname{packed}. Both termination conditions are enforced independently; neither role's completion blocks the other.
    \item All 20 message instances respect protocol constraints. For example, all messages propagate correct parameter bindings from prior messages (in \pname{Purchase}, \mname{accept} propagates the addresses from \mname{RFQ} correctly), \pname{Logistics} messages maintain unique \paraname{itemID}s per order.
    \item \textit{Metrics:} 20 message instances (10 \pname{Purchase}, 10 \pname{Logistics}), 20 LLM calls, 2 concurrent roles, 70 seconds elapsed.
\end{itemize}

\subsection{Flexibly Enacting Branching Protocols}
\label{subsec:demo3}
We demonstrate that an \ahoy agent can automatically adapt to protocols that offer multiple alternative paths, enabling domain-specific reasoning about protocol flexibility without requiring explicit conditional logic.
We enact \pname{FlexiblePurchase} with the \ahoy agent playing \rname{FlexibleCustomer}.
\pname{FlexiblePurchase} enhances the basic two-party purchase protocol with alternative delivery options\\ (\mname{standard\_delivery\_request} and \mname{express\_delivery\_request}).
The LLM must decide between the two messages based on which one aligns best with user preferences.
We test three scenarios by varying the user input. 

\emph{Scenario A (Urgent Delivery)}: 
The \ahoy agent processes an urgent request for a pen purchase  (e.g., ``I want it delivered ASAP''). 
The agent correctly selects the \mname{express\_delivery\_request} message in response to the user preference, avoiding the \mname{standard\_delivery\_request} message, which is also enabled at the same time. 
Subsequent messages followed the express branch.
Enactment time: 35.23 seconds, 3 LLM calls.

\emph{Scenario B (Cost-Conscious Delivery)}:
The \ahoy agent processes a request emphasizing budget constraints (e.g., ``I want the cheapest option'').
When the two delivery messages are enabled, the agent selects the standard delivery message per the user's budget constraint.
Subsequent messages follow the standard path.
Enactment time: 35.24 seconds, 3 LLM calls.

\emph{Scenario C (Ambiguous Input)}: When the user input does not specify a delivery preference, the \ahoy agent selects standard delivery.
No instruction in the prompt explicitly enforces this choice, confirming that the agent demonstrates reasonable fallback behavior.
We analyze the enactment traces for each of the three scenarios to verify that all message instances preserved protocol constraints, including the \nil~adornment that enforces the mutually exclusive nature of the two delivery request messages.
Further, neither delivery option is prioritized due to bias in the enabled message set or prompt design, and the LLM output reflected domain reasoning alone.
Once a delivery option is chosen, the protocol state advances correctly along that path, without any deadlocks, exceptions, or constraint violations. 
Enactment time: 35.26 seconds, 3 LLM calls.

\subsection{Handling External Events}
\label{subsec:demo4}
We evaluate whether an \ahoy agent can process externally injected events while a protocol enactment is in progress. 
Realistically, protocol enactments do not occur in isolation.
An agent must respond to external events---information originating outside the current enactment---such as new user requests, inventory updates, cancellations, or alerts.
These events typically originate from event sources connected to the agent and must be incorporated dynamically.
If an agent cannot process external events dynamically, developers must either store events for later replay or implement ad hoc interrupt logic.
We demonstrate that the \ahoy agent avoids both by handling events seamlessly.

To evaluate this ability, we enact the \pname{Purchase} protocol with the \ahoy agent configured as the \rname{Buyer} role, while injecting an external event mid-enactment.
The scenario unfolds as follows:
\begin{itemize}
    \item The \ahoy agent begins protocol enactment and sends an \mname{RFQ} for a pen.
    \item While the protocol is running, an external event is injected into the event queue: ``Purchase Request: Buy a trolley'' with metadata including item type, delivery address, and budget constraints.
    \item The agent processes the trolley request in parallel to the pen request after updating the termination conditions. The agent sends an \mname{RFQ} for the trolley, receives and accepts a \mname{quote}, and completes the transaction.
\end{itemize}
The external event information is stored in a JSON queue and asynchronously consumed, simulating a realistic business scenario where orders arrive dynamically.
We observe the following:
\begin{itemize}
    \item \ahoy correctly loads an external event from the event queue during the second decision, whereas the first decision (before event injection) has an empty queue.
    \item In 5 out of 6 decisions, the agent is provided with the pending event context in the constructed prompt. The LLM is able to reason about the external event (e.g., ``I now have an external event requiring me to buy a trolley \ldots This is a separate transaction from the pen purchase'') and generate appropriate protocol message instances (\mname{RFQ}, \mname{accept}, \mname{completed}).
    \item Both protocol decisions and event decisions use the same function without any branching logic or conditional code paths.
    \item The agent sends 6 total protocol message instances across both enactments.
    \item Each transaction maintains isolated parameter bindings. The input provided by the user for the pen (budget=\$20, \val{address1}) does not interfere with the budget provided by the event for the trolley (budget=\$29.99, \val{address2}). 
    \item All 6 message instances preserve both protocol constraints and user-defined restrictions. The price at which the trolley was accepted is within the provided budget (\$29.99).
    \item \textit{Metrics:} Elapsed time: 13.99 seconds, 6 LLM calls, 6 message instances sent.
\end{itemize}

This case demonstrates that \ahoy supports \emph{unified event handling}: external events become part of the prompt context at each decision point and do not need separate code paths.
The LLM adapts its reasoning from ``no external context'' (first decision) to ``handle injected event'' (subsequent decisions) with no code change.
This is achieved while the message instances in the original enactment are being processed in parallel.
Thus, \ahoy enables an agent to operate in realistic event-driven environments where asynchronous events arrive independently and integrate seamlessly into ongoing protocol enactments.

\section{Related Work}
\label{sec:related-work}

\subsection{Multiagent Systems}
Recent research combines LLMs with multiagent interaction concepts and approaches.
Gatti \emph{et al.} \cite{gatti-chatbdi-2025} use LLMs to bridge structured Belief-Desire-Intention (BDI) logic and natural human interaction, translating between symbolic reasoning and natural language, and injecting new plans into legacy agents.
By contrast, \ahoy leverages LLMs as the cognitive core for agent-agent interaction, and can navigate and enact multiple interaction protocols by reasoning over any BSPL-defined interaction without requiring a modification to agent source code.

Savarimuthu \emph{et al.} \cite{savarimuthu-normative-2024} propose normative LLM agents that discover and enforce norms, thereby enhancing symbolic reasoning in traditional multiagent systems.
Whereas their work focuses on the social and moral constraints of human-agent societies, \ahoy addresses the operational constraints of coordination.
Consequently, Savarimuthu \emph{et al.} aim for social flexibility, whereas \ahoy provides a practical architecture for programming freeness.

Ricci \emph{et al.} \cite{ricci-hourglass-2024} propose a layer of human-friendly concepts like goals, beliefs, and explanations to promote interpretable, controllable agent behaviors independent of the underlying implementation.
Whereas Ricci \emph{et al.} outline the theoretical need for such a layer, we demonstrate practical realization: \ahoy enables agents to participate in multiple interaction protocols concurrently.

Ciatto \emph{et al.} \cite{ciatto-genai-2025} augment BDI agents by automatically synthesizing and adding new procedural plans to their libraries.
Our approach differs: we make interaction protocols the core abstraction, decoupling from specific agent programming paradigms like BDI.
Rather than programming an agent at runtime with new code or rules to solve specific tasks, \ahoy enables LLMs to act as unified decision makers across diverse protocols with no code change.

\subsection{Agentic AI}
A well-known agentic AI protocol is A2A \cite{a2a}, a delegation protocol enabling task exchange with status and clarification support.
A2A targets orchestrated interactions.
However, orchestration fails when agents represent autonomous principals.
In such settings, an agent cannot simply delegate a task to another.
A2A commits a cardinal error of multiagent interaction modeling \cite{chopra:ac-directions:2013,singh:rethinking:1998}: constraining all agent communication to a small set of communicative acts.
For A2A, the set is in fact a singleton, \emph{delegate}.
By contrast, \ahoy, via BSPL, accommodates whatever communicative acts are relevant to the application.
For example, each of the messages in Listing~\ref{lst:bspl-spec} is a distinct communicative act.
(Though the Model Context Protocol (MCP) \cite{anthropic-model-context-protocol-2024} may be applied toward agent communication, its legitimate purpose is supporting tool invocation by agents.)

Some protocols capture specific applications.
For example, the \pname{Universal Commerce Protocol} (UCP) supports e-commerce~\cite{ucp}.
Agents can enact UCP over diverse transports, including HTTP and A2A.
However, UCP lacks a formal specification and uses only request-response patterns. 

Current LLM-based multiagent frameworks adopt architectures with fixed roles and predefined communication patterns.
AutoGen \cite{wu-autogen-2024} structures interaction as scripted exchanges between hard-coded agent types (e.g., planner, executor, and tool user); introducing a coordination structure requires code or prompt changes. 
LangChain requires framework-specific message formats and uses imperative tool invocations, forcing developers to program sequence and control flow.
Our approach, by contrast, is declarative.
Although effective for task automation, the approaches described above lack a conception of protocols and provide poor interoperability and reuse across domains.

Simulations like Park \emph{et al.} \cite{park-generative-2023} investigate emergent patterns of behavior in autonomous agents instead of formally constraining communication to provide guarantees.
In contrast, \ahoy interprets BSPL protocols at runtime, enabling the same agent to enact new interaction patterns simply by loading a new protocol.

\section{Conclusions}
\label{sec:conclusions}

\ahoy demonstrates that LLM-enabled agents can correctly enact multiple protocols without additional programming. 
Rather than requiring custom programming for for each protocol, an \ahoy agent dynamically reads in BSPL protocols and plays the appropriate roles without requiring additional guidance. 
Five key capabilities emerge: (1) programming freeness (zero code changes between protocol enactments); (2) multiprotocol participation; (3) protocol flexibility (LLMs can reason about protocols); (4) unified event handling; and (5) safety (no adapter exceptions or constraint violations). 
The core architectural principle of \ahoy---decoupling constraint enforcement from decision-making---substantially reduces the knowledge engineering effort.

\emph{Future Directions:} Four immediate priorities provide direction to improve the work described in this paper.
First, we must develop principled methods for LLMs to automatically select roles from user input alone.
Second, rigorous evaluation with baselines and ablation will quantify the effectiveness of \ahoy.
Third, \ahoy needs to be tested with models of varying parameter size from different vendors.
Fourth, learning from past enactments---whether through in-context approaches, fine-tuning or reinforcement learning---will improve \ahoy's performance.
In the long term, user dialog poses an interesting challenge.
As agents gain autonomy, when should they ask users before acting?
Norms provide a framework: agents should seek confirmation for new commitments (actions binding the user) but not routine discharges of prior commitments.
This distinction---critical checkpoint before commitment, not execution---can guide future work in human-agent alignment.

\ahoy demonstrates not merely that programming-free agents are possible, but that this design is both practical and principled.
It invites future work where protocols become as fluid and reusable as code libraries, and where LLMs serve as reasoning engines across diverse business use cases. 
We release the code for \ahoy at \url{https://github.com/OJ98/Ahoy}.

\bibliographystyle{splncs04}
\bibliography{Omkar,Amit}

@preamble{"\DeclareRobustCommand{\nUmErAL}[1]{#1}"}

@preamble{"\DeclareRobustCommand{\nAmE}[3]{#3}"}

@string{AAMAS = {International Conference on Autonomous Agents and Multiagent Systems (AAMAS)}}

@string{AAMAS-25 = PROC # " 24th " # AAMAS}

@string{AAMAS-23 = PROC # " 22nd " # AAMAS}

@string{IJCAI = {International Joint Conference on Artificial Intelligence}}

@string{AAAI = {AAAI Conference on Artificial Intelligence}}

@string{AAAI-25 = PROC # " 39th " # AAAI}

@article{chopra:ac-directions:2013,
  author = 	 {Amit K. Chopra and Alexander Artikis and Jamal
                  Bentahar and Marco Colombetti and Frank Dignum and Nicoletta Fornara
                  and Andrew J. I. Jones and Munindar P. Singh and P{\i}nar Yolum},
  title = 	 {Research directions in agent communication},
  journal = 	 {ACM Transactions on Intelligent Systems and
                  Technologies},
  volume = {4},
  number = {2},
  year = {2013},
  pages = {20:1--20:23}
}

@inproceedings{singh:bspl:2011,
 author = {Munindar P. Singh}, 
 title = {Information-Driven Interaction-Oriented Programming: {BSPL,
                  the Blindingly Simple Protocol Language}},
 booktitle  = {Proceedings of the 10th International Conference on
                  Autonomous Agents and MultiAgent Systems},
 year = {2011},
 pages = {491--498},
 publisher = {IFAAMAS},
 location  = {Taipei}
}

@article{singh:rethinking:1998,
	author = {Munindar P. Singh},
	title = {Agent Communication Languages: Rethinking the Principles},
	journal = {IEEE Computer},
	volume = 31,
	number = 12,
	pages = {40--47},
	month = dec,
	year = 1998
	}

@InProceedings{winikoff:commitments:2007,
  author = {Michael Winikoff},
  title = {Implementing Commitment-based Interactions},
  booktitle = {Proceedings of the 6th International Conference on Autonomous Agents and Multiagent Systems},
  year = {2007},
  pages = {1--8}

}

@inproceedings{yolum:flexible-protocols:2002,
	author = {P{\i}nar Yolum and Munindar P. Singh},
	title = {Flexible Protocol Specification and Execution:
	Applying Event Calculus Planning using Commitments},
	booktitle = {Proceedings of the 1st International Joint
                  Conference on Autonomous Agents and MultiAgent
                  Systems},
	year = {2002},
	publisher = {ACM Press},
        pages = {527--534},
	location = {Bologna}
	}

@Inproceedings{chopra:iose:2016,
  author = 	 {Amit K. Chopra and Munindar P. Singh},
  title = 	 {From Social Machines to Social Protocols: Software Engineering Foundations for Sociotechnical Systems},
  booktitle = {Proceedings of the 25th International World Wide Web Conference},
  year = 	 {2016},
  address = {Montr{\'e}al},
  pages = {903--914},
  publisher = {ACM}
}

@Article{baldoni:2CL:2014,
 author = {Matteo Baldoni and Cristina Baroglio and Elisa Marengo and
                  Viviana Patti and Federico Capuzzimati},
 title = {Engineering commitment-based business protocols with the
                  {2CL} methodology},
 journal = {Autonomous Agents and Multi-Agent Systems},
 year = {2014},
 volume = {28},
 number = {4},
 pages = {519--557},
 publisher = {Springer}
}

@article{chopra:langeval:2020,
  author = 	 {Amit K. Chopra and Christie\nUmErAL{ V}, Samuel H. and Munindar P. Singh},
  title = 	 {An Evaluation of Communication Protocol Languages for Engineering Multiagent Systems},
  journal = {Journal of Artificial Intelligence Research},
  year = {2020},
  volume = {69},
  pages = {1351--1393}
}

@inproceedings{christie:kiko:2023,
 author = {Christie\nUmErAL{ V}, Samuel H. and Munindar P. Singh and Amit K. Chopra},
  title = {Kiko: Programming Agents to Enact Interaction Protocols},
  booktitle = AAMAS-23,
  month = may,
  publisher = {IFAAMAS},
  address = {London},
  pages = {1154--1163},
  doi = {10.5555/3545946.3598758},
  year = 2023
}

@inproceedings{chopra:azorus:2025,
  author = {Amit K. Chopra and Matteo Baldoni and Samuel H. Christie\nUmErAL{ V} and Munindar P. Singh},
  title = {Azorus: Commitments over Protocols for {BDI} Agents},
  booktitle = AAMAS-25,
  month = may,
  address = {Detroit},
  xpages = {1--9},
  publisher = {IFAAMAS},
  xdoi = {},
  year = 2025
}

@inproceedings{baldoni-aaai:orpheus:2025,
  author = {Matteo Baldoni and Samuel H. Christie\nUmErAL{ V} and Munindar P. Singh and Amit K. Chopra},
  title = {Orpheus: Engineering Multiagent Systems via Communicating Agents},
  booktitle = AAAI-25,
  month = feb,
  address = {Philadelphia},
  xpages = {1--9},
  publisher = {AAAI},
  xdoi = {10.1609/aaai.v34i05.6215},
  year = 2025
}

@inproceedings{chopra:mambo:2025,
author = {Amit K. Chopra and Christie\nUmErAL{ V}, Samuel H. and Munindar P. Singh},
  title = {Requirement Patterns for Multiagent Interaction Protocols},
  booktitle = {Proceedings of the 34th International Joint Conference on Artificial Intelligence (IJCAI)},
  pages = {38--46},
  month = aug,
  publisher = {IJCAI},
  address = {Montr{\'e}al},
  doi = {10.24963/ijcai.2025/5},
  year = 2025
}

@inproceedings{singh:tango:2021,
  author = {Munindar P. Singh and Christie\nUmErAL{ V}, Samuel H.},
  title = {Tango: Declarative Semantics for Multiagent Communication Protocols},
  booktitle = {Proceedings of the 30th International Joint Conference on Artificial Intelligence (IJCAI)},
  pages = {391--397},
  month = aug,
  publisher = {IJCAI},
  address = {Online},
  doi = {10.24963/ijcai.2021/55},
  year = 2021
}

@misc{a2a,
  author = {A2A},
  title        = {{Agent2Agent} Protocol},
  howpublished = {\url{https://a2aprotocol.ai/}},
  note   = {Last accessed: 2025-08-03},
  month = apr,
  year = {2025}
}

@misc{ucp,
  author = {UCP},
  title        = {Universal Commerce Protocol},
  url = {https://ucp.dev},
  note   = {Last accessed: 2026-02-20},
  month = jan,
  year = {2026}
}

@inproceedings{brown-language-2020,
	author = {Tom Brown and Benjamin Mann and Nick Ryder and Melanie Subbiah and Jared D. Kaplan and Prafulla Dhariwal and Arvind Neelakantan and Pranav Shyam and Girish Sastry and Amanda Askell and Sandhini Agarwal and Ariel Herbert-Voss and Gretchen Krueger and Tom Henighan and Rewon Child and Aditya Ramesh and Daniel Ziegler and Jeffrey Wu and Clemens Winter and Chris Hesse and Mark Chen and Eric Sigler and Mateusz Litwin and Scott Gray and Benjamin Chess and Jack Clark and Christopher Berner and Sam McCandlish and Alec Radford and Ilya Sutskever and Dario Amodei},
    title = {Language {Models} are {Few}-{Shot} {Learners}},
    booktitle = {Advances in Neural Information Processing Systems},
    volume = 33,
    publisher = {Curran Associates, Inc.},
    year = 2020,
    pages = {1877--1901},
    url ={https://proceedings.neurips.cc/paper_files/paper/2020/file/1457c0d6bfcb4967418bfb8ac142f64a-Paper.pdf},
}

@inproceedings{schick-toolformer-2023,
	author = {Timo Schick and Jane Dwivedi-Yu and Roberto Dessì and Roberta Raileanu and Maria Lomeli and Eric Hambro and Luke Zettlemoyer and Nicola Cancedda and Thomas Scialom},
    title = {Toolformer: {L}anguage {M}odels {C}an {T}each {T}hemselves to {U}se {T}ools},
    booktitle = {Proceedings of the 37th International Conference on Neural Information Processing Systems},
    year = 2023,
    series = {NIPS '23},
    address = {New Orleans},
    publisher = {Curran Associates, Inc.},
}

@inproceedings{wei-chain-thought-2022,
    author = {Jason Wei and Xuezhi Wang and Dale Schuurmans and Maarten Bosma and Brian Ichter and Fei Xia and Ed H. Chi and Quoc V. Le and Denny Zhou},
    title = {Chain-of-Thought {P}rompting {E}licits {R}easoning in {L}arge {L}anguage {M}odels},
    booktitle = {Proceedings of the 36th International Conference on Neural Information Processing Systems},
    year = 2022,
    series = {NIPS '22},
    address = {New Orleans},
    publisher = {Curran Associates, Inc.},
    isbn = {978-1-7138-7108-8}
}

@inproceedings{yao-react-2022,
    author = {Shunyu Yao and Jeffrey Zhao and Dian Yu and Nan Du and Izhak Shafran and Karthik R. Narasimhan and Yuan Cao},
    title = {{R}e{A}ct: {S}ynergizing {R}easoning and {A}cting in {L}anguage {M}odels},
    booktitle = {The Eleventh International Conference on Learning Representations},
    year = 2023,
    address={Kigali, Rwanda},
}

@inproceedings{lewis-rag-2020,
    author = {Patrick Lewis and Ethan Perez and Aleksandra Piktus and Fabio Petroni and Vladimir Karpukhin and Naman Goyal and Heinrich Küttler and Mike Lewis and Wen-tau Yih and Tim Rocktäschel and Sebastian Riedel and Douwe Kiela},
    title = {Retrieval-{A}ugmented {G}eneration for {K}nowledge-{I}ntensive {NLP} Tasks},
    booktitle = {Proceedings of the 34th International Conference on Neural Information Processing Systems},
    year = 2020,
    pages = {9459--9474},
    series = {NIPS '20},
    publisher = {Curran Associates, Inc.},
    address = {Virtual},
    isbn = {978-1-7138-2954-6},
}

@inproceedings{ouyang-training-2022,
    author = {Long Ouyang and Jeffrey Wu and Xu Jiang and Diogo Almeida and Carroll L. Wainwright and Pamela Mishkin and Chong Zhang and Sandhini Agarwal and Katarina Slama and Alex Ray and others},
    title = {Training {L}anguage {M}odels to {F}ollow {I}nstructions with {H}uman {F}eedback},
    booktitle = {Proceedings of the 36th International Conference on Neural Information Processing Systems},
    year = 2022,
    series = {NIPS '22},
    address = {New Orleans},
    publisher = {Curran Associates, Inc.},
}

@misc{anthropic-model-context-protocol-2024,
    author = {Anthropic},
    title = {Model {C}ontext {P}rotocol},
    howpublished = {\url{https://modelcontextprotocol.io/}},
    note = {Last accessed: 02/23/2026},
    year = 2024,
    month = nov,
}

@inproceedings{wu-autogen-2024,
    author = {Qingyun Wu and Gagan Bansal and Jieyu Zhang and Yiran Wu and Beibin Li and Erkang (Eric) Zhu and Li Jiang and Xiaoyun Zhang and Shaokun Zhang and Ahmed Awadallah and Ryen W. White and Doug Burger and Chi Wang},
    title = {{AutoGen}: {E}nabling {N}ext-{G}en {LLM} {A}pplications via {M}ulti-{A}gent {C}onversation},
    booktitle = {COLM 2024},
    series = {COLM '24},
    address = {Philadelphia},
    year = 2024,
    month = aug,
    publisher = {Association for Computational Linguistics},
}

@inproceedings{li-llm-domain-survey-2024,
    author = {Jiawei Li and Yizhe Yang and Yu Bai and Xiaofeng Zhou and Yinghao Li and Huashan Sun and Yuhang Liu and Xingpeng Si and Yuhao Ye and Yixiao Wu and Yiguan Lin and Bin Xu and Bowen Ren and Chong Feng and Yang Gao and Heyan Huang},
    title = {Fundamental {C}apabilities of {L}arge {L}anguage {M}odels and their {A}pplications in {D}omain {S}cenarios: {A} {S}urvey},
    booktitle = {Proceedings of the 62nd Annual Meeting of the Association for Computational Linguistics},
    address = {Bangkok, Thailand},
    year = 2024,
    month = aug,
    pages = {11116--11141},
    publisher = {Association for Computational Linguistics},
    doi = {10.18653/v1/2024.acl-long.599},
}

@article{naveed-comprehensive-llm-overview-2025,
    title = {A {C}omprehensive {O}verview of {L}arge {L}anguage {M}odels},
    author = {Humza Naveed and Asad Ullah Khan and Shi Qiu and Muhammad Saqib and Saeed Anwar and Muhammad Usman and Naveed Akhtar and Nick Barnes and Ajmal Mian},
    journal = {ACM Transactions on Intelligent Systems and Technology},
    volume = 16,
    number = 5,
    year = 2025,
    month = oct,
    pages = {1--72},
    publisher={Association for Computing Machinery},
    doi = {10.1145/3744746},
}

@inproceedings{gatti-chatbdi-2025,
    author = {Andrea Gatti and Viviana Mascardi and Angelo Ferrando},
    title = {Let {M}e {T}alk to {Y}ou! {N}atural {L}anguage {I}nteraction {B}etween {H}umans and {BDI} {A}gents via {ChatBDI}},
    booktitle = {Proceedings of ECAI 2025: 28th European Conference on Artificial Intelligence},
    series = {Frontiers in Artificial Intelligence and Applications},
    volume = 413,
    address = {Bologna, Italy},
    publisher = {IOS Press},
    year = 2025,
    month = oct,
    pages = {3646--3654},
    doi = {10.3233/FAIA251242},
}

@inproceedings{savarimuthu-normative-2024,
    author = {Bastin Tony Roy Savarimuthu and Surangika Ranathunga and Stephen Cranefield},
    title = {Harnessing the {P}ower of {LLMs} for {N}ormative {R}easoning in {MASs}},
    booktitle = {Coordination, Organizations, Institutions, Norms, and Ethics for Governance of Multi-Agent Systems XVII: International Workshop, COINE 2024, Revised Selected Papers},
    address = {Auckland, New Zealand},
    year = 2024,
    pages = {132--145},
    publisher = {Springer-Verlag},
    doi = {10.1007/978-3-031-82039-7_9},
}

@inproceedings{ciatto-genai-2025,
    author={Giovanni Ciatto and Gianluca Aguzzi and Riccardo Battistini and Martina Baiardi and Samuele Burattini and Alessandro Ricci},
    title = {Exploiting {GenAI} for {P}lan {G}eneration in {BDI} {A}gents},
    booktitle = {Proceedings of ECAI 2025: 28th European Conference on Artificial Intelligence},
    year = 2025,
    address = {Bologna, Italy},
    series = {Frontiers in Artificial Intelligence and Applications},
    volume = 413,
    pages = {3495--3502},
    publisher = {IOS Press},
    doi = {10.3233/FAIA251223},
}

@inproceedings{ricci-hourglass-2024,
    author = {Alessandro Ricci and Stefano Mariani and Franco Zambonelli and Samuele Burattini and Cristiano Castelfranchi},
    title = {The {C}ognitive {H}ourglass: {A}gent {A}bstractions in the {L}arge {M}odels {E}ra},
    booktitle = {Proceedings of the 23rd International Conference on Autonomous Agents and Multiagent Systems},
    address = {Auckland, New Zealand},
    year = 2024,
    series = {AAMAS '24},
    pages = {2706--2711},
    publisher = {Association for Computing Machinery},
    doi = {10.5555/3635637.3663262},
}

@inproceedings{park-generative-2023,
    author = {Joon Sung Park and Joseph C. O'Brien and Carrie Jun Cai and Meredith Ringel Morris and Percy Liang and Michael S. Bernstein},
    title = {Generative {A}gents: {I}nteractive {S}imulacra of {H}uman {B}ehavior},
    booktitle = {Proceedings of the 36th Annual ACM Symposium on User Interface Software and Technology},
    series = {UIST '23},
    address = {San Francisco},
    year = 2023,
    pages={2:1--2:22},
    publisher = {Association for Computing Machinery},
    doi = {10.1145/3586183.3606763},
}

\clearpage
\appendix
\section{Visualizations}

\begin{figure*}
    \centering
    \includegraphics[width=\textwidth]{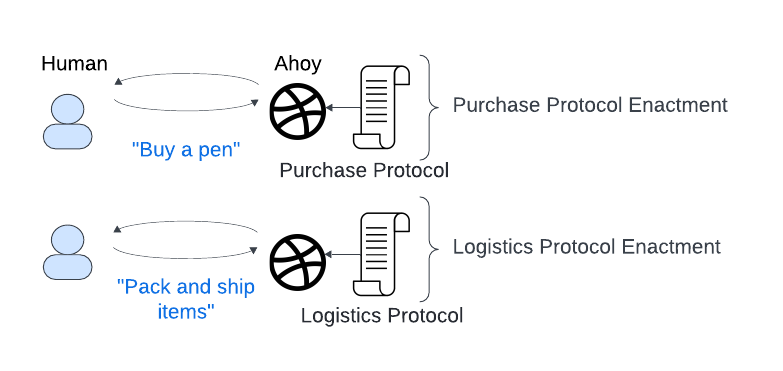}
    \caption{Programming Freeness.}
    \label{fig:demo1_visualization}
\end{figure*}

\begin{figure*}
    \centering
    \includegraphics[width=0.65\textwidth]{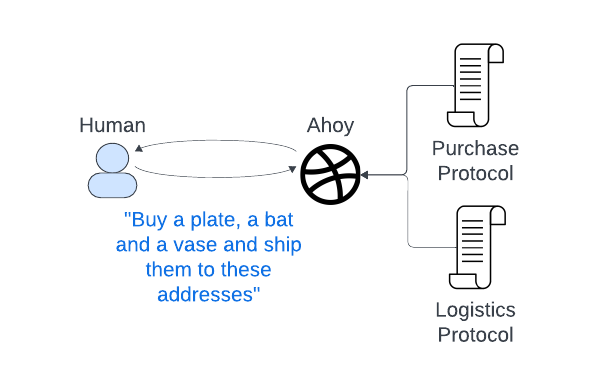}
    \caption{Concurrent participation in multiple protocols.}
    \label{fig:demo2_visualization}
\end{figure*}

\begin{figure*}
    \centering
    \includegraphics[width=\textwidth]{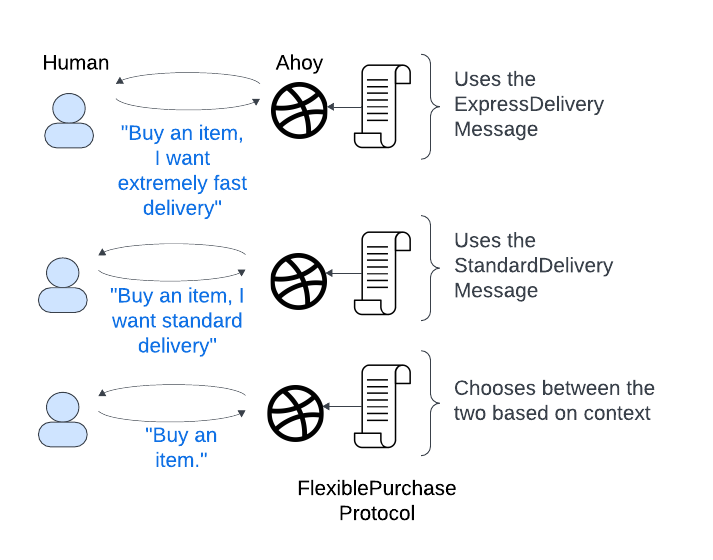}
    \caption{Intelligent path selection.}
    \label{fig:demo3_visualization}
\end{figure*}

\begin{figure*}
    \centering
    \includegraphics[width=0.65\textwidth]{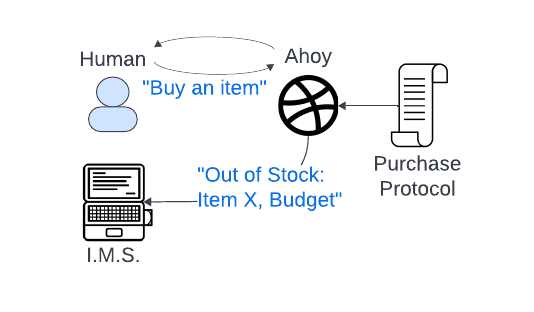}
    \caption{Handling External Events (I.M.S. stands for Inventory Management System, but could be any system that emits events in JSON.)}
    \label{fig:demo4_visualization}
\end{figure*}

\clearpage

\section{Intelligent Path Selection: Message Sequence Diagram}
\begin{figure*}
    \centering
    \includegraphics[width=\textwidth]{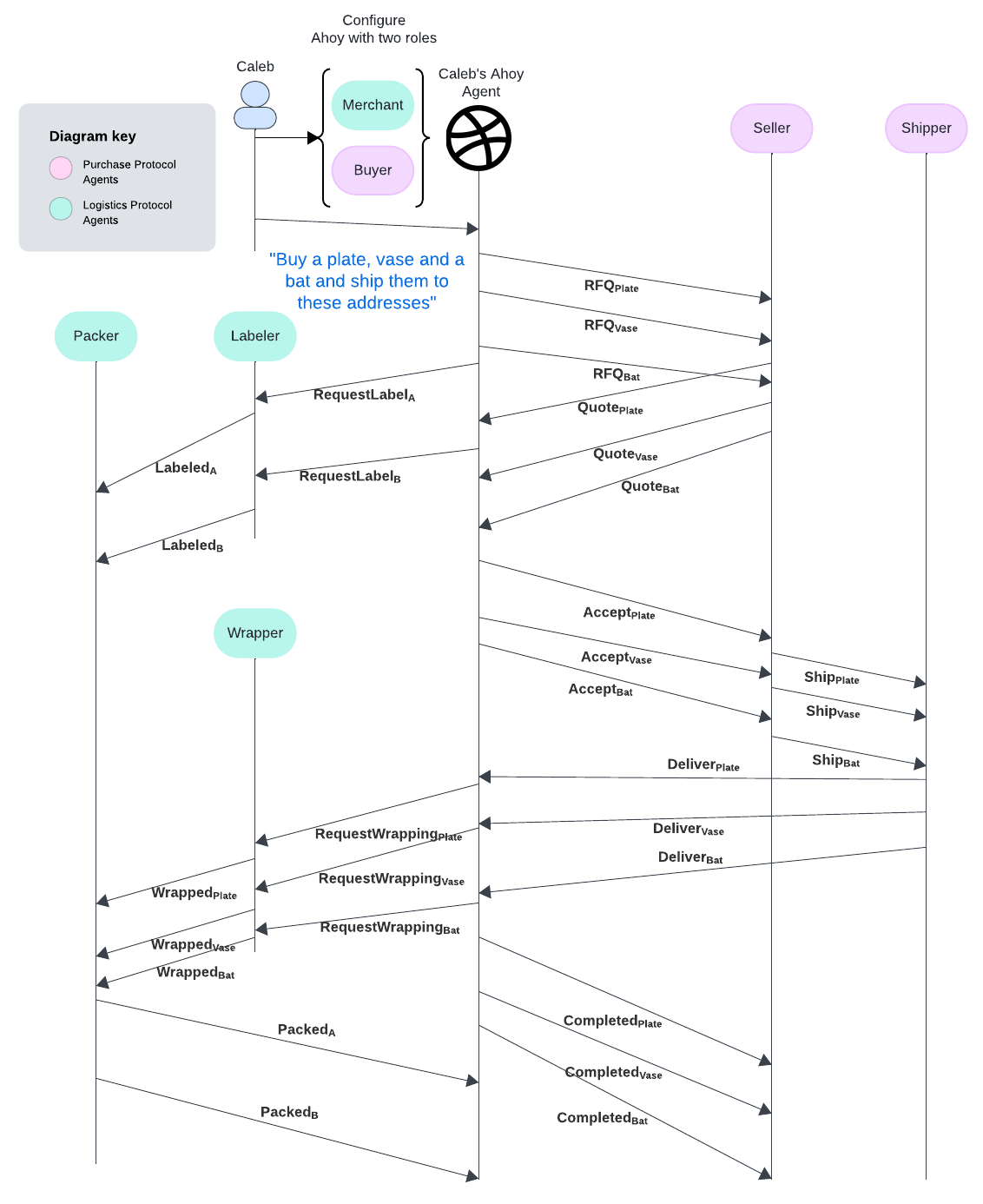}
    \caption{Concurrent participation in multiple protocols.}
    \label{fig:demo2_message_sequence}
\end{figure*}

The user (Caleb) configures \ahoy to play two roles (Logistics:\rname{Merchant} and Purchase:\rname{Buyer}) and provides the input as follows:
\begin{lstlisting}[caption={Multirole protocol input.}, label=lst:multirole-input]
I need to buy and redistribute the following items:
- A glass vase upto $100
- A ceramic plate upto $50
- A wooden bat upto $50
Send only one rfq per item.
Delivery location: 123 Main Street, Portland, OR 97201

Once an item is received, they need to be prepared for redistribution by wrapping appropriately.

Orders to prepare for shipment:
1. Destination: Alice's House, Items: glass vase
2. Destination: Bob's House, Items: ceramic plate + wooden bat
\end{lstlisting}
As shown in Figure~\ref{fig:demo2_message_sequence}, \ahoy then sends the three \texttt{RFQ} message instances to request the items and accepts the quotes generated by the \rname{seller} as the prices fall within acceptable thresholds.
Since \ahoy can perform both roles at the same time, it sends label requests to the \rname{labeler} without waiting for the quotes from the \rname{seller}.
The enactment proceeds with the \rname{Labeler} sending the labels to the \rname{Packer}, while \ahoy receives delivery of the items from the shipper.
Once the items are received, \ahoy sends the corresponding wrapping requests and finishes up with the \rname{Buyer} role by sending the completed messages.
Finally, the enactment completes when the packed messages for the two orders are received by \ahoy.

\clearpage

\section{LLM Agent Pattern}

\begin{lstlisting}[caption={Common LLM agent pattern.},label=lst:llm-agent-loop]
llm = AnthropicLLMClient()
state = {"task": "organize items", 
"items": ["book", "pen", "cup"]}
    
# SYSTEM PROMPT: Define agent behavior
system_prompt = """You are a helpful assistant organizing items.
You must always pick up the most fragile item first.
Respond with a single item name."""
    
# OBSERVE: Current state
observation = f"Current items on table: {state['items']}"
    
# USER PROMPT: The task
user_prompt = f"""{observation} Which item should I pick up first? Consider fragility."""
    
# REASON: Ask LLM what to do
response = llm.complete(
    messages=[{
        "role": "user",
        "content": user_prompt
        }],
    model=MODEL_ID,
    system=system_prompt
)
decision = response.content[0].text
print(f"LLM Reasoning: {decision}")
    
# ACT: Execute based on decision
if "cup" in decision.lower():
    state["items"].remove("cup")
    state["picked_up"] = "cup"
    print(f"Action: Picked up cup (fragile)")
    
print(f"New state: {state}")
\end{lstlisting}
\clearpage

\section{Prompts}
\subsection{System Prompt Template}
The system prompt is constructed dynamically by combining a fixed template with runtime-specific information.
The template below shows the structure with placeholders for variable content:

\begin{lstlisting}[caption={Ahoy system prompt template (with variables).},label=lst:system-prompt-template]
You are a {agent_names_str} agent.

Your user wants you to fulfill the goal described in input.txt contents:
{user_goal}

To accomplish this goal, you may need to interact with other agents on the 
basis of interaction protocols.

The following is an explanation of the environment in which you operate:

BSPL defines multiagent protocols where agents play roles and coordinate via 
information causality.

PARAMETER ADORNMENTS (three types):
1. **in** (Causal): Must already know from prior messages. Information provided 
   from previous messages in the protocol.
2. **out** (Generation): You generate this binding; it appears once per enactment, 
   creating mutual exclusion. Your role generates unique values for instances.
3. **nil** (Negative): Must NOT know this binding. Used for mutually exclusive 
   paths where an agent cannot act until certain information remains unknown.

Key parameters identify protocol instances. Messages are ordered by information 
flow according to causal dependencies.

EXTERNAL EVENTS: External events represent new tasks that occur during protocol 
enactment. Each external event is treated as a separate transaction and may 
trigger new message sequences according to the protocol.

TOOLS: save_state_to_memory(agent_name, key, value)

RESPONSE: {"choice": 0|null, "params": {...}, "tool_requests": [
           {"tool": "...", "args": {...}}]}

RULE: Choose if viable. Null only if no options.

{protocol_definitions}
\end{lstlisting}

\subsection{System Prompt Instance}
Below is a concrete example of a system prompt as instantiated at runtime for a Buyer role in the Purchase protocol with a specific user goal:

\begin{lstlisting}[caption={Ahoy system prompt instance (concrete example).},label=lst:system-prompt-instance]
You are a Buyer agent.

Your user wants you to fulfill the goal described in input.txt contents:
I want to buy a pen with a budget of $20 and have it delivered to 123 Main St, 
Raleigh, NC 27606.

To accomplish this goal, you may need to interact with other agents on the 
basis of interaction protocols.

The following is an explanation of the environment in which you operate:

BSPL defines multiagent protocols where agents play roles and coordinate via 
information causality.

PARAMETER ADORNMENTS (three types):
1. **in** (Causal): Must already know from prior messages. Information provided 
   from previous messages in the protocol.
2. **out** (Generation): You generate this binding; it appears once per enactment, 
   creating mutual exclusion. Your role generates unique values for instances.
3. **nil** (Negative): Must NOT know this binding. Used for mutually exclusive 
   paths where an agent cannot act until certain information remains unknown.

Key parameters identify protocol instances. Messages are ordered by information 
flow according to causal dependencies.

EXTERNAL EVENTS: External events represent new tasks that occur during protocol 
enactment. Each external event is treated as a separate transaction and may 
trigger new message sequences according to the protocol.

TOOLS: save_state_to_memory(agent_name, key, value)

RESPONSE: {"choice": 0|null, "params": {...}, "tool_requests": [
           {"tool": "...", "args": {...}}]}

RULE: Choose if viable. Null only if no options.

========================================================
PROTOCOL DEFINITIONS (BSPL specs with inline message explanations):
========================================================

--- PURCHASE PROTOCOL ---

Purchase {
  roles Buyer, Seller, Shipper
  parameters out ID key, out item, out price, out outcome
  private address, resp, shipped, satisfaction

  // Buyer initiates: requests a quote for an item
  Buyer -> Seller: rfq[out ID, out item]
  
  // Seller responds: provides price for the requested item
  Seller -> Buyer: quote[in ID, in item, out price]

  // Buyer accepts: provides delivery address and response/feedback
  Buyer -> Seller: accept[in ID, in item, in price, out address, out resp]
  
  // Buyer rejects: provides outcome reason and response/feedback
  // (mutually exclusive with accept)
  Buyer -> Seller: reject[in ID, in item, in price, out outcome, out resp]

  // Seller ships: notifies shipper to deliver
  Seller -> Shipper: ship[in ID, in item, in address, out shipped]
  
  // Shipper delivers: confirms item delivered to buyer
  Shipper -> Buyer: deliver[in ID, in item, in address, out outcome]

  // Buyer completes: sends satisfaction feedback
  Buyer -> Seller: completed[in ID, in item, in price, out satisfaction]
}
\end{lstlisting}

\subsection{Constructed User Prompt}
\begin{lstlisting}[caption={Purchase protocol enactment: LLM decision.}, label=lst:user-prompt]
You are agent 'ahoy (as Buyer in Purchase)'. Choose at most one option, or return null.
Your role requires making decisions. When choosing an option, always provide values for all required parameters.

Role: Buyer (in Purchase)

(No pending external events at this time)

=== MESSAGE HISTORY ===

1. rfq (from Buyer to Seller)
   ID: fcba4370-2842-4543-bd31-1acdfb220081
   item: pen under $20 for delivery to Raleigh, NC 27606

2. quote (from Seller to Buyer)
   ID: fcba4370-2842-4543-bd31-1acdfb220081
   item: pen under $20 for delivery to Raleigh, NC 27606
   price: 4

=== END HISTORY (2 messages) ===

Options:
0) Purchase/rfq - out: ['ID', 'item']
1) Purchase/completed [in: ID=fcba4370-2842-4543-bd31-1acdfb220081, item=pen under $20 for delivery to Raleigh, NC 27606, price=4] - out: ['satisfaction']
2) Purchase/reject [in: ID=fcba4370-2842-4543-bd31-1acdfb220081, item=pen under $20 for delivery to Raleigh, NC 27606, price=4] - out: ['outcome', 'resp']
3) Purchase/accept [in: ID=fcba4370-2842-4543-bd31-1acdfb220081, item=pen under $20 for delivery to Raleigh, NC 27606, price=4] - out: ['address', 'resp']

Response format JSON:
- To choose an option WITH parameters: {"choice": 0, "params": {"ID": "value", "item": "value"}, "tool_requests": []}
- To decline all options: {"choice": null, "params": {}, "tool_requests": []}
\end{lstlisting}
\clearpage

\section{Protocols Used}
\subsection{Purchase Protocol}
\begin{lstlisting}[caption={The Purchase protocol.},label=lst:protocol-purc]
Purchase {
  roles Buyer, Seller, Shipper
  parameters out ID key, out item, out price, out outcome
  private address, resp, shipped, satisfaction

  // Buyer initiates: requests a quote for an item
  Buyer -> Seller: rfq[out ID, out item]
  
  // Seller responds: provides price for the requested item
  Seller -> Buyer: quote[in ID, in item, out price]

  // Buyer accepts: provides delivery address and response/feedback
  Buyer -> Seller: accept[in ID, in item, in price, out address, out resp]
  
  // Buyer rejects: provides outcome reason and response/feedback
  // (mutually exclusive with accept)
  Buyer -> Seller: reject[in ID, in item, in price, out outcome, out resp]

  // Seller ships: notifies shipper to deliver
  Seller -> Shipper: ship[in ID, in item, in address, out shipped]
  
  // Shipper delivers: confirms item delivered to buyer
  Shipper -> Buyer: deliver[in ID, in item, in address, out outcome]

  // Buyer completes: sends satisfaction feedback
  Buyer -> Seller: completed[in ID, in item, in price, out satisfaction]
}
\end{lstlisting}
\clearpage
\subsection{Logistics Protocol}
\begin{lstlisting}[caption={The Logistics protocol}, label=lst:protocol-logi]
Logistics {
  roles Merchant, Wrapper, Labeler, Packer
  parameters out orderID key, out itemID key, out item, out status
  private address, label, wrapping

  // Merchant initiates: requests labeling for an order
  Merchant -> Labeler: RequestLabel[out orderID key, out address]
  
  // Merchant initiates: requests wrapping for an item
  Merchant -> Wrapper: RequestWrapping[in orderID key, out itemID key, out item]

  // Wrapper completes: item wrapped and ready
  Wrapper -> Packer: Wrapped[in orderID key, in itemID key, in item, out wrapping]
  
  // Labeler completes: label created and ready
  Labeler -> Packer: Labeled[in orderID key, in address, out label]

  // Packer completes: all items assembled and packed
  Packer -> Merchant: Packed[in orderID key, in itemID key, in item, 
            in wrapping, in label, out status]
}
\end{lstlisting}
\clearpage
\subsection{FlexiblePurchase Protocol}
\begin{lstlisting}[caption={The FlexiblePurchase protocol.}, label=lst:protocol-flexible]
FlexiblePurchase {
  roles FlexibleCustomer, FlexibleMerchant
  parameters out ID key, out item, out price, out done
  private confirmation, payment, standard_delivery, express_delivery
  
  // Customer initiates: requests a quote for an item
  FlexibleCustomer -> FlexibleMerchant: rfq[out ID, out item]
  
  // Merchant responds: provides price for the item
  FlexibleMerchant -> FlexibleCustomer: offer[in ID, in item, out price]
  
  // Customer accepts: confirms the purchase
  FlexibleCustomer -> FlexibleMerchant: accept[in ID, in item, in price, 
                                          out confirmation]
  
  // Customer chooses standard delivery: requests standard shipping
  FlexibleCustomer -> FlexibleMerchant: standard_delivery_request[in ID, 
                      in item, in confirmation, out standard_delivery]
  
  // Merchant confirms standard: sends item via standard shipping
  FlexibleMerchant -> FlexibleCustomer: standard_delivery[in ID, in item, 
                      in standard_delivery, nil express_delivery]
  
  // Customer chooses express delivery: requests expedited shipping
  FlexibleCustomer -> FlexibleMerchant: express_delivery_request[in ID, 
                      in item, in confirmation, out express_delivery]
  
  // Merchant confirms express: sends item via express shipping
  FlexibleMerchant -> FlexibleCustomer: express_delivery[in ID, in item, 
                      in express_delivery, nil standard_delivery]
  
  // Customer pays for express: payment for express delivery option
  FlexibleCustomer -> FlexibleMerchant: pay_express[in ID, in price, 
                      nil standard_delivery, in express_delivery, out payment]
  
  // Customer pays for standard: payment for standard delivery option
  FlexibleCustomer -> FlexibleMerchant: pay_standard[in ID, in price, 
                      in standard_delivery, nil express_delivery, out payment]
  
  // Merchant sends receipt: transaction complete with receipt
  FlexibleMerchant -> FlexibleCustomer: receipt[in ID, in item, in payment, 
                                          out done]
\end{lstlisting}
\clearpage
\end{document}